\documentclass[twocolumn,english,showpacs,showkeys,reprint,linenumbers]{revtex4}
\usepackage[T1]{fontenc}
\usepackage[utf8]{inputenc}
\usepackage{amsmath}
\usepackage{amssymb}
\usepackage{graphicx}

\makeatletter
\@ifundefined{textcolor}{}
{%
 \definecolor{BLACK}{gray}{0}
 \definecolor{WHITE}{gray}{1}
 \definecolor{RED}{rgb}{1,0,0}
 \definecolor{GREEN}{rgb}{0,1,0}
 \definecolor{BLUE}{rgb}{0,0,1}
 \definecolor{CYAN}{cmyk}{1,0,0,0}
 \definecolor{MAGENTA}{cmyk}{0,1,0,0}
 \definecolor{YELLOW}{cmyk}{0,0,1,0}
}

\usepackage{babel}

\usepackage{babel}

\usepackage{babel}

\usepackage{babel}

\usepackage{babel}

\makeatother

\usepackage{babel}
\begin{document}

\title{Magnetization reversal behavior in complex shaped Co nanowires: a
nanomagnet morphology optimization}

\author{Fatih Zighem }

\email{zighem@univ-paris13.fr}

\author{Silvana Mercone}

\email{mercone@univ-paris13.fr}

\affiliation{Laboratoire des Sciences des Procédés et des Matériaux, CNRS-Université
Paris XIII, Sorbonne Paris Cité, Villetaneuse, France}

\date{October 22$^{nd}$, 2014 }
\begin{abstract}
A systematic micromagnetic study of the morphological characteristic
effects over the magnetic static properties of Co-based complex shaped
nanowires is presented. The relevance of each characteristic size
(i.e. length $L$, diameter $d$, and size of the nanowires head $T$)
and their critical values are discussed in the coercive field optimization
goal. Our results strongly confirms that once the aspect ratio ($\frac{L}{d}$)
of the nanowire is bigger than around 10, the length is no more the
pertinent parameter and instead the internal diameter and the shape
of the nanowires play a key role. We attribute this behavior to the
non uniform distribution of the demagnetizing field which is localized
in the nanowires head and acts as a nucleation point for the incoherent
magnetization reversal. Finally, angular dependence of the magnetization
are simulated and compared to the case of a prolate spheroid for all
considered morphologies. 
\end{abstract}

\keywords{micromagnetic simulations, magnetic nano-objects, magnetization curves}

\maketitle

\section{Introduction}

Magnetic one-dimensional nanowires have shown peculiar properties
interesting for a bunch of different possible applications. Among
them it can be mentioned the high density magnetic recording\cite{Weller1999},
the microwave devices\cite{Piraux1999} and also a new generation
of permanent magnets\cite{Maurer_APL,Gutfleisch2011}. In this latter
case the nanowire can constitute the nano-magnet itself inside the
surrounding media or it can also be the building block of dense nanostructured
magnet bulk systems. In both cases the key issue is to maximize the
magnet performances at a nanoscale level (i.e. high remanent magnetization
and large coercive field). The most common approach to solve this
issue, is to use non-spherical nanoparticles as they may naturally
display a large shape anisotropy \cite{Cowburn2000,Bance} added to
the magnetocrystalline one and thus easily beat the unwished superparamagnetic
behavior usually occurring for nanosized particles at room temperature
\cite{Weller1999}. The basic idea is then to maximize the effective
magnetic anisotropies (both shape and magnetocrystalline ones) in
ferromagnetic systems (i.e. Co and Fe based for instance). In this
context, coercive fields above 10 kOe and remanent to saturation magnetization
ratio near 1 are expected in single crystal cobalt and iron based
nanorods. Several ways have been developed for the elaboration of
such nano-objects: i) templates methods which involve the growth inside
a host nanoporous matrix of nanorods or nanowires usually by electrochemical
deposition \cite{sellmyer2001,whitney,Pan2005,zhang_2003} and ii)
alternative chemical ones involving their synthesis in a solution.
The soft chemistry route presents several advantages compared to the
electrochemical one \cite{Dumestre2002,Liu2010,D.Ung_2005,D.Ung2007,Seung2005}:
the low coast price, the relatively low synthesis temperature, the
high crystalline quality and pureness of the chemical phase and also
the good dispersion of the nanowires. On the other hand, despite this
good control, the nanowires obtained by the chemical route usually
show complex shaped one-dimensional morphology \cite{Liu2010,Soumare2009_Co,Viau2009}
and have never shown the expected theoretical coercive field value
\cite{Maurer_APL,Respaud2009}. Previous theoretical studies tried
to explain the observed gap by focusing on how the effective magnetic
anisotropy can be mainly decreased both by complex shape of the nanowires
and by a non-single crystal structure. At its turn, this change in
the effective anisotropy modify significantly the magnetization reversal
processes and consequently affects the magnetic properties wished
for applications \cite{Vivas2012,Sokalski2011,Ott_JAP}. Those previous
results highlighted, on the one hand, the importance of crystal structure
uniformity (i.e. absence of stacking faults) in order not to locally
decrease the magnetocrystalline anisotropy coefficient along the nanowire.
On the other hand, they also shown that different morphologies can
induce very unwished magnetic coercivity meltdown. This latter has
been attributed to the presence of a demagnetizing field distribution
at the edge of the nanowires which seems to be strictly linked to
their complex shape and to let the reversal mechanism happening at
lower coercive field. Unexpectedly, the demagnetizing values were
found to be very high both in non-complex shaped nanowires (i.e. cylinder
type) presenting a high coercive behavior then in complex shaped ones
having a very modest coercivity. All those previous studies put in
evidence the still open question of the understanding of the magnetization
reversal mechanism in nanowires systems as well as the presence of
concomitant behaviors that can act contradictorily for the optimization
of the nanowires magnetic properties \cite{S=00003D00003D0000E1nchez2009}.
For example, it is easy to understand that a long nanowire giving
rise to a high aspect ratio (length of nanowire versus diameter in
the middle of the nanowire) is needed for the optimization of the
shape anisotropy coefficient while a very long nanowire increases
dramatically the probability of a high density of stacking faults
(low uniformity of the nanostructure) and consequently improve the
probability of decreasing the magnetocrystalline effective coefficient.
Thus, if the general idea coming up from a vast landscape of experimental
and theoretical studies, is that a long nanowire will improve the
effective magnetic anisotropy, it also goes in the other way. Thus,
the issue concerning the compromise to reach between the contradictory
effects coming from structure and morphology optimization is still
relevant.

Thereby, a systematic study of the morphological characteristics effects
over the magnetic static properties of Co-based complex shaped nanowires
is proposed here. Starting from the experimental shapes observed,
we performed a micromagnetic study of the reversal magnetization mechanism
varying continuously the nanowire geometry. The relevance of each
characteristic size (i.e. length, diameter and size of the head) and
their critical values are discussed in the coercive field optimization
goal.

\section{Micromagnetic Method}

\begin{figure}
\includegraphics[bb=25bp 440bp 340bp 585bp,clip,width=8.5cm]{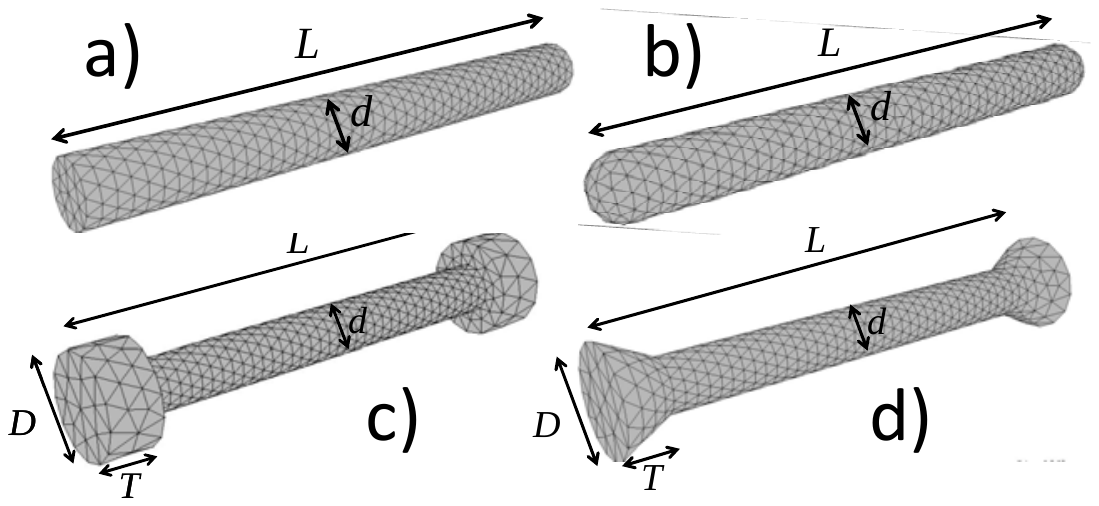}

\protect\protect\protect\caption{Typical meshes of the different considered nano-objects: a) cylinder,
b) capped cylinder, c) dumbbell and d) diabolo). The length of these
nano-objects is $L=100$ nm while the internal diameter is $d=10$
nm. Moreover, the parameters $D$ and $T$ are equal to 20 nm and
10 nm, respectively. Note that the distance between two nodes is smaller
then 4 nm in all the simulated meshes.}

\label{Fig_Nano-objects} 
\end{figure}

The micromagnetic calculations have been performed by using the \textsc{nmag}
micromagnetic simulation package \cite{fishbacher2007}. This software
is indeed well suited to describe the complex shape of the nano-objects
fabricated thanks to the polyol process as it allows to simulate very
different shapes related to the one-dimensional simple nanowire. The
magnetic parameters of all the studied nano-objects correspond to
typical values of cobalt material\cite{tannewald}. Thus the saturation
magnetization ($M_{S}=1.4\times10^{3}$ emu.cm$^{-3}$) and the exchange
stiffness ($A=1.2\times10^{-6}$ erg.cm$^{-1}$) have been defined
from the bulk reference. In order to dissociate the effect of the
morphology from the one coming from the structure, we considered a
single crystal nanowire (i.e. the magnetocrystalline anisotropy is
kept constant and in our case equal to zero). In this condition, an
exchange length ($\ell_{ex}=\sqrt{A/2\pi M_{S}^{2}}$) around $3.2$
nm is then deduced. Once defined the magnetic properties, an important
step of the procedure consists in defining the objects geometries
and to their discretization by using a mesher (\textsc{netgen} and
\textsc{gmsh} in the present case \cite{netgen,gmsh}). Figure \ref{Fig_Nano-objects}
presents the typical nano-objects which will be considered throughout
this work, namely: a) a cylinder, b) a capped cylinder, c) a dumbbell
and c) a diabolo. Thereafter, $z$-direction (see Figure \ref{Fig_Nano-objects})
will corresponds to the revolution axis of the studied objects (considered
parallel to the length of the nanowires). All these nano-objects are
characterized by a length $L$ and a diameter in the middle $d$.
Two more geometrical parameters are necessary for the diabolo and
the dumbbell morphology study: the tip width $D$ and the tip thickness
$T$. We compared the results obtained for these specific nano-objects
to the well-known standard geometry of a prolate spheroid (ellipsoid)
\cite{Sun2005}. The length $L$ of our nanowires varies from 50 nm
to 2000 nm while the internal diameter vary from 6 to 20 nm. Typical
length and diameters experimentally observed by using the polyol process
elaboration are within these ranges. It is important to underline
here that the distance between two nodes of the different meshes used
in all the simulated geometries, is most of the time smaller than
$\ell_{ex}$. Finally, the \textsc{nmag} package is employed to solve
the Landau–Lifshtiz–Gilbert equation. As only the static magnetization
configurations are examined in all the performed simulations, the
damping Gilbert constant ($\alpha$) has been set to 0.5 in order
to minimize the computing time. The magnetization reversal mechanism
of the different nanowires has been studied mainly throughout the
magnetization curves calculations along $z$ direction. The magnetization
curves are investigated by applying an external magnetic field $\vec{H}$
ranging from $-10$ kOe to $+10$ kOe; it is incremented by steps
of about 50 Oe.

\section{Results and discussion}

\begin{figure}
\includegraphics[bb=35bp 140bp 410bp 580bp,clip,width=8.5cm]{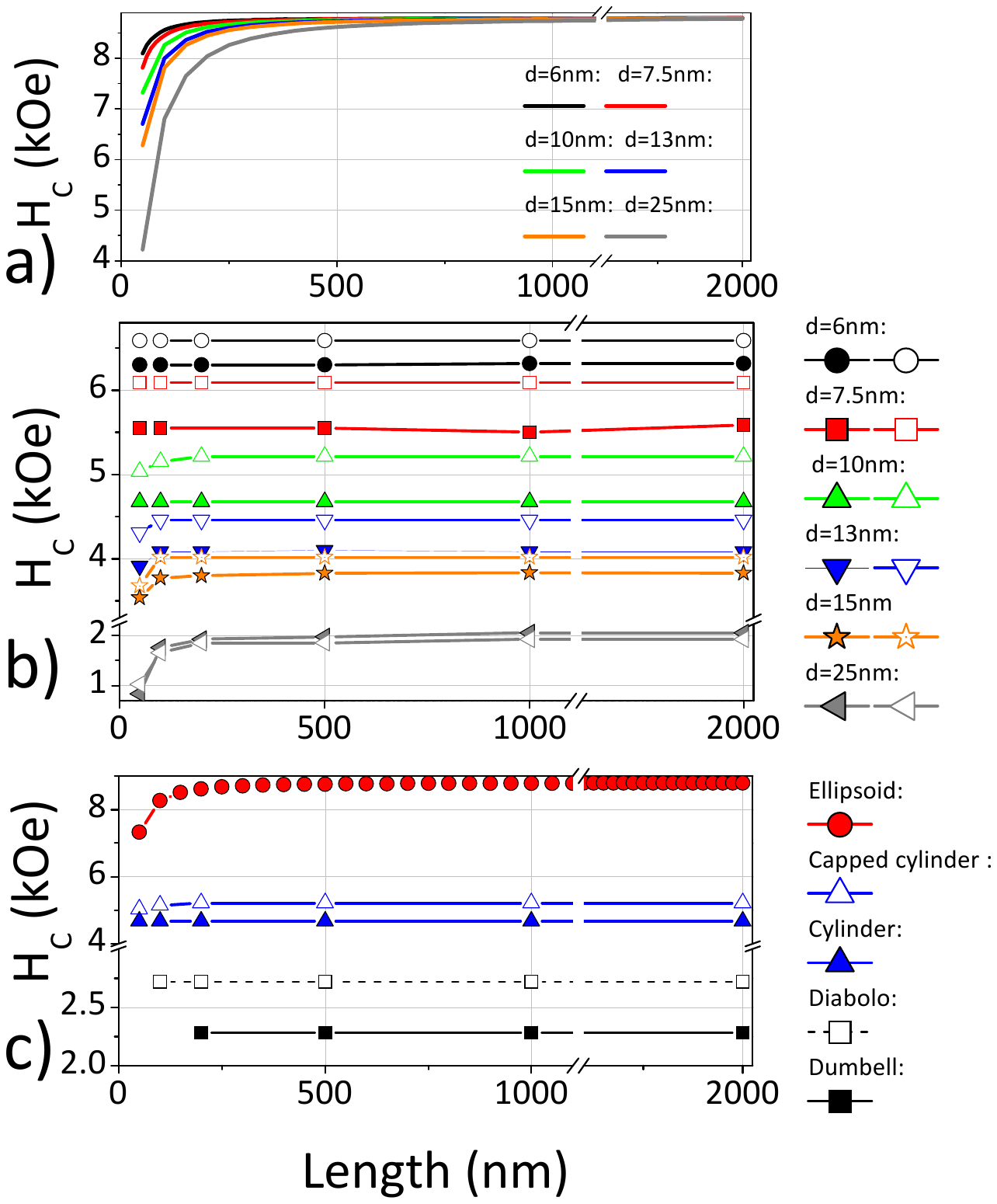}

\protect\protect\protect\caption{Coercive field along the revolution axis as function of the length
$L$ for: a) prolate spheroid objects with different aspect ratios
b) cylinder (filled symbols) and capped cylinder (open symbols) nanowires.
c) Comparison of the coercive field among the various shaped objects
as function of the length $L$ and for a diameter in the middle $d=10$
nm. The parameters $D$ and $T$ for the dumbbell and the diabolo
nanowires have been fixed to 20 nm and 10 nm respectively.}

\label{Fig_Length_Dependence} 
\end{figure}

As previously mentioned, one of the still open question that we addressed
here, is the range of the length size of the nanowires playing a role
on the static magnetic properties of these latter. As already detailed
in the introduction, there is an important compromise to reach between
a long nanowire and a perfect crystallized one. In this purpose we
address here the question of the minimum limit for the nanowire length,
necessary for a given geometry to optimize the shape anisotropy coefficient.
The ellipsoidal (prolate spheroid) nano-objects will be used as reference
because their well known static magnetic behavior set the upper limit
of the shape anisotropy coefficient for a given length ($L$) and
internal diameter ($d$)\cite{Bance}. In this latter standard geometry,
two of the three axis have the same length and the third one is bigger
than the two others. In this configuration, the coercive field $H_{C}$
along the revolution axis is simply given by \cite{Sun2005}: 
\begin{equation}
H_{C}=4\pi\left(N_{z}-N_{x}\right)\label{eq:HC_SW}
\end{equation}
 where $N_{z}$ and $N_{x}$ (and $N_{y}$) are the demagnetizing
factors which can be analytically determined by using the following
relations \cite{Sun2005}: 
\begin{equation}
N_{x}+N_{y}+N_{z}=1
\end{equation}

\begin{equation}
N_{z}=\frac{1}{a^{2}-1}\left(\frac{a}{2\left(a^{2}-1\right)^{0.5}}\log\left(\frac{a+\left(a^{2}-1\right)^{0.5}}{a-\left(a^{2}-1\right)^{0.5}}\right)-1\right)\label{eq:Nz_factor}
\end{equation}

where $a$ is the aspect ratio ($a=L/d$). Figure \ref{Fig_Length_Dependence}a)
presents the calculated coercive field of ellipsoidal nano-object
for various diameters ($d$) ranging from 6 to 25 nm and as function
of $L$ (i. e. aspect ratio varying from 2 to 330) by using equations
\ref{eq:HC_SW} and \ref{eq:Nz_factor}. It clearly appears that $H_{C}$
tends to the value of an infinite cylinder ($H_{C}=2\pi M_{S}=8800$
Oe) for all the diameter when $L$ increases. This tendency, as expected,
is more and more efficient as the length over internal diameter ratio
increases. At small aspect ratios (below $L=500$ nm), a non-negligible
$H_{C}$-variation is observed, especially for the higher considered
diameters. For instance, for $d=25$ nm (resp. $d=6$ nm), the coercive
field varies from $4.2$ kOe (resp. 8 kOe) to 8.6 kOe (resp. 8.8 kOe)
for a length varying from 50 to 500 nm. The observed behavior points
out that whatever is the size of $L$ and $d$, once the aspect ratio
is bigger than around 10 the coercive field reaches the maximum. This
is confirmed also by all the others shaped nano-objects (see Figure
2c)). Thus, in case of one-dimensional nano-objects for typical diameter
$d$ around 10 nm, a minimum length of 100 nm is sufficient to reach
the maximum coercive field value. All the experimental effort focusing
on improving this aspect ratio value by extending the length of nanowires
at the fixed $d$, will be useless in term of shape anisotropy optimization
\cite{Respaud2009}. This is true whatever is the morphology of the
nanowires. Another interesting feature in Figure \ref{Fig_Length_Dependence}b)
is that the coercive field variations calculated by micromagnetic
simulations on capped cylinders and cylinders nanowires of similar
aspect ratios have a different behavior. The most important difference
is that the calculated $H_{C}$ values strongly depend on the diameter
$d$ while it remains constant in almost the whole $L$ range. Thus
in a large range of length ($L=100-2000$ nm) the smaller is the internal
diameter $d$, the higher is the coercive field. This result strongly
confirms that once the aspect ratio ($\frac{L}{d}$) of the nanowire
is bigger then 10, the length is no more the pertinent parameter and
instead the internal diameter and the shape (cylinder or capped cylinder)
of the nanowires play a key role. The shape effect is strongly confirmed
by the diabolos and dumbbells nano-objects behavior presented in Figure
\ref{Fig_Length_Dependence}c). In this graph, a comparison in between
all the four discussed morphologies is performed for an internal diameter
of 10 nm. The parameters $D$ and $T$ of the dumbbells and the diabolos
tips have been fixed to 20 nm and 10 nm, respectively. Impressively,
nanowires with the same aspect ratio, show very different coercive
behavior. This confirms the importance of the edge morphology over
the magnetic shape anisotropy efficiency. In order to analyze this
effect, we focused our attention on the demagnetizing field distribution
inside the nanowires.

\begin{figure}
\includegraphics[bb=20bp 430bp 660bp 595bp,clip,width=8.5cm]{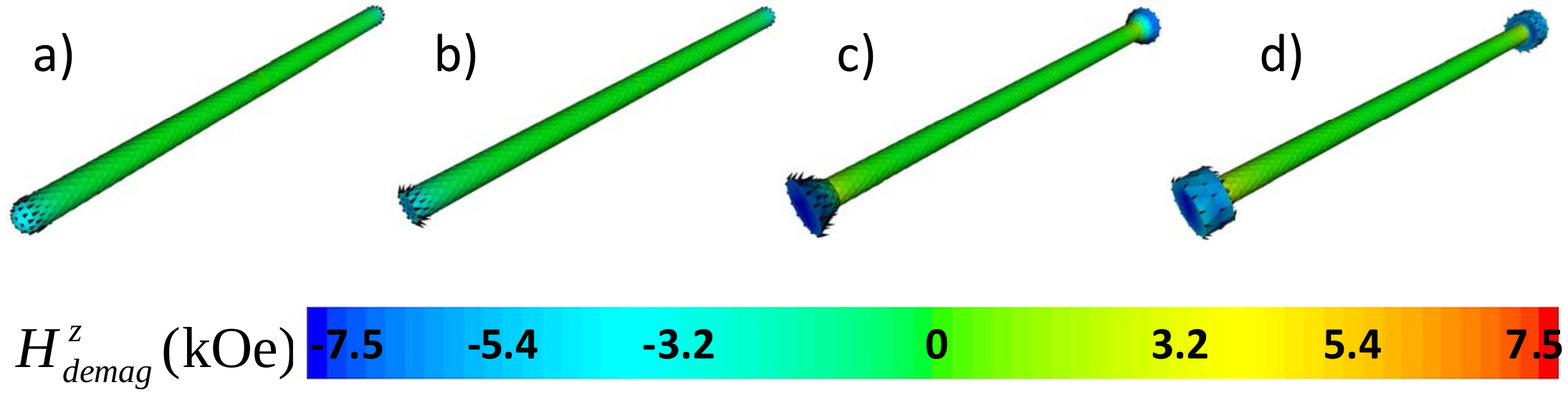}

\protect\protect\protect\caption{Demagnetizing field distributions inside the different nano-objects
at a saturating field: a) capped cylinder, b) cylinder, c) dumbbell
and d) diabolo). The geometrical parameters are fixed as follows:
$L=200$ nm; $d=10$ nm ($D=20$ nm and $T=10$ nm for the dumbbell
and the diabolo). The colors encode the $z$-component of the demagnetizing
field $H_{dem}$ ($x$- and $y$-components are close to zero in all
the presented cases).}

\label{Fig_Demag_3D_Saturate} 
\end{figure}

As previously reported \cite{Bance,Ott_JAP}, for a prolate spheroid
(ellipsoidal nano-object) the demagnetizing field is uniform and equal
to $\vec{H}_{dem}=-4\pi\overset{=}{N}\vec{M}$. Figure \ref{Fig_Demag_3D_Saturate}
shows the demagnetizing field distributions (along the revolution
axis) at a saturating magnetic field applied (i.e. the magnetic moments
are aligned along this $z$ axis). These distributions have been calculated
for nano-objects of 200 nm in length and 10 nm in diameter (i.e. $\frac{L}{d}>10$).
It is worth mentioning that demagnetizing field distributions and
nucleation volume variations during the reversal process have been
largely studied in literature in individual magnetic bits of different
geometry (prism, sphere and cylinder one) (i.e. $\frac{L}{d}<10$)
and thus will be not reconsidered here\cite{Bance,krone}. Also it
should be underlined that in our case, because of the uniform distributions
of the magnetic moments, only the $z$-component of $\vec{H}_{dem}$
are non zero (see color map of Figure \ref{Fig_Demag_3D_Saturate}).
More precisely, the demagnetizing field is almost zero in the whole
nano-objects except at the tips where values in the range 4.0-7.5
kOe can be reached. The higher values are obtained for the diabolo
and dumbbell shapes (around 7.5 kOe for the maximum value) whereas
smaller values are obtained in the capped cylinder and cylinder shapes
(around 4 kOe for the maximum value). Similar $\vec{H}_{dem}$-distributions
are obtained whatever the value of the length $L$ (above 100 nm)
which again means that once the aspect ratio is fixed, these distributions
are the key parameter in the nucleation of the magnetization reversal
mechanism. In order to study into details this shape effect, we decided
to compare the hysteresis loops of the four different shapes discussed.

\begin{figure}
\includegraphics[bb=60bp 10bp 460bp 590bp,clip,width=8.5cm]{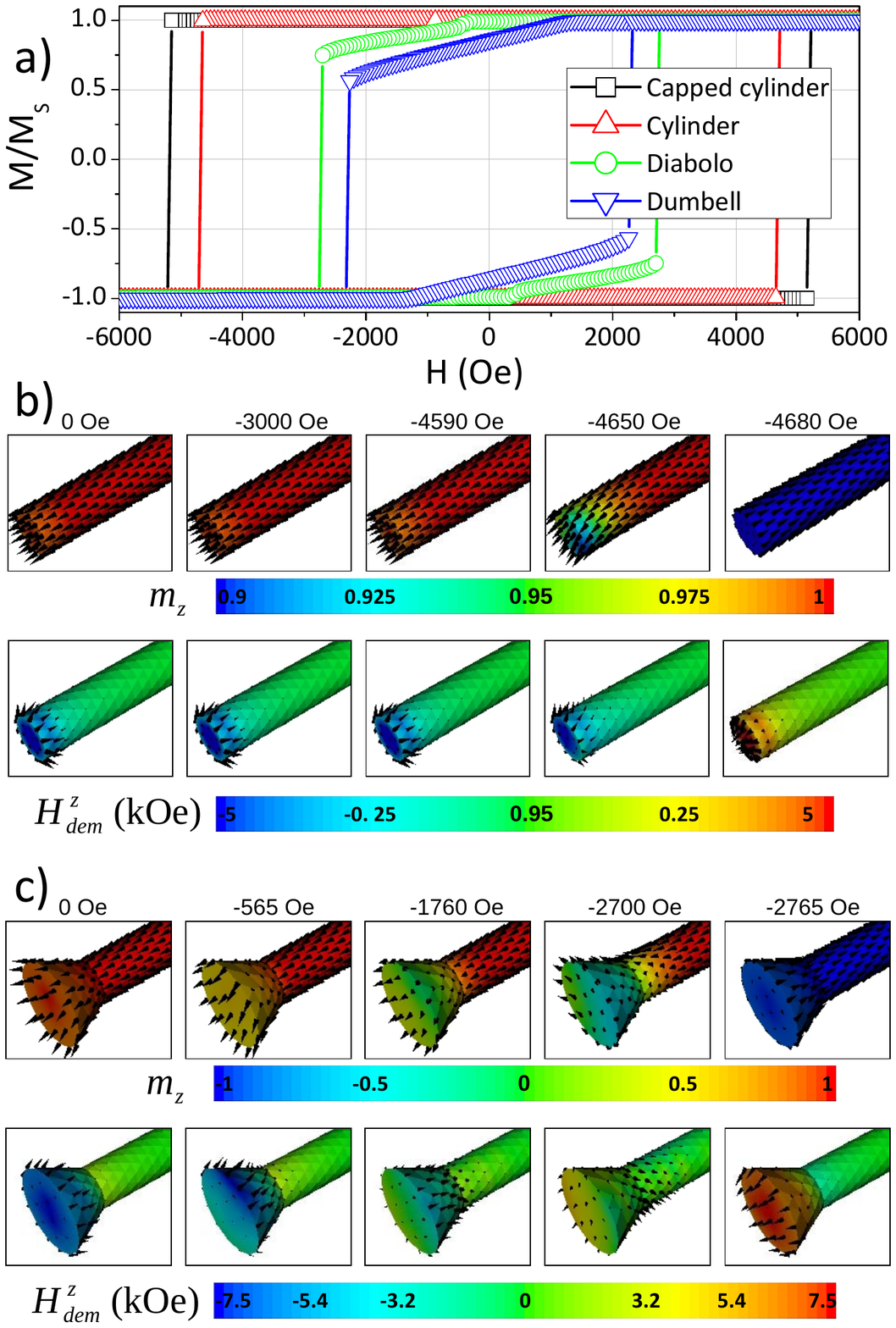}

\protect\protect\caption{a) Magnetization curves calculated along the revolution axis for the
different types of nano-objects. b)c) Magnetization (up images) and
demagnetizing field (down images) distributions inside the cylinder
(b) and the diabolo (c) during the reversal process. The geometrical
parameters are fixed as follows: $L=200$ nm; $d=10$ nm ($D=20$
nm and $T=10$ nm for the dumbbell and the diabolo)}

\label{Fig_Diabolos_Reversal} 
\end{figure}

Looking at Figure \ref{Fig_Diabolos_Reversal}a), the magnetization
curves calculated along the revolution axis for the different types
of nano-objects, we can separate them into two groups. Indeed, two
hysteresis loops shapes are recognized: on the one hand the cylinder
and capped cylinder nano-objects hysteresis behavior and on the other
hand the dumbbell and diabolo nano-objects ones. While the hysteresis
loops of the cylinders and capped cylinders are relatively squares
(for applied field along the revolution axis), the ones of diabolos
and dumbbells clearly presents a linear dependence before the reversal
which is characteristics of a non uniform magnetization distribution.
Figures \ref{Fig_Diabolos_Reversal}b) and c) present the magnetization
distributions and the corresponding demagnetizing field distributions
inside the cylinder and the diabolo during the reversal process. The
distributions of the capped cylinder and the dumbbell objects are
not shown here, as their configurations have been found close to the
ones of the cylinder and the diabolo, respectively. The magnetization
distribution is almost uniform during the whole reversal process for
the cylinder whereas it presents non-negligible $x-$ and $y-$components
at the tips during the reversal mechanism in the diabolo case. The
linear decreasing dependencies of $\frac{M_{R}}{M_{S}}$ ratios during
the reversal mechanism for the second group of nano-objects, are clearly
due to these non uniform magnetization distribution. In this case
the magnetization remains uniform only before ($H<500$ Oe) and after
($H\sim-2800$ Oe) the reversal process. Comparing the demagnetizing
field of these two groups of nano-objects in the magnetic state where
the magnetization is uniform, it is interesting to note that lowest
coercive fields are obtained for nano-objects presenting the higher
localized demagnetizing field (i. e. dumbbell and diabolo). In both
cases demagnetizing field is localized at the tip and is very high
while the magnetization is uniform (i.e. before and after the reversal).
During the reversal mechanism (see Figure \ref{Fig_Diabolos_Reversal}c)),
it is clear that the demagnetizing field can reach very low values
(close to zero) even for the second group of nano-objects. All those
results bring to the conclusion that the tip play the role of nucleation
point for the reversal of the magnetization. Figure \ref{Fig_Diabolos_Reversal}a)
also shows that the $\frac{M_{R}}{M_{S}}$ ratios are lower for the
dumbbell and diabolo nanowires, as already stressed out. This ratio
is dependent of the volume of the nano-objects, so it is possible
to increase them by increasing the length of the nano-objects. For
this shaped nano-objects then, increasing the $\frac{M_{R}}{M_{S}}$
ratios would be the only reason for an elaboration effort towards
longer nano-objects while either for the coercivity optimization as
stated before and for thermal stability improvement\cite{krone},
this effort would be useless. In order to go into details of the nucleation
influence of the demagnetizing field on the magnetization reversal
mechanism, we studied the coercive behavior of the dumbbell and diabolo
nano-objects as function of the thickness $T$ of the edge.

\begin{figure}
\includegraphics[bb=55bp 220bp 690bp 585bp,clip,width=8.5cm]{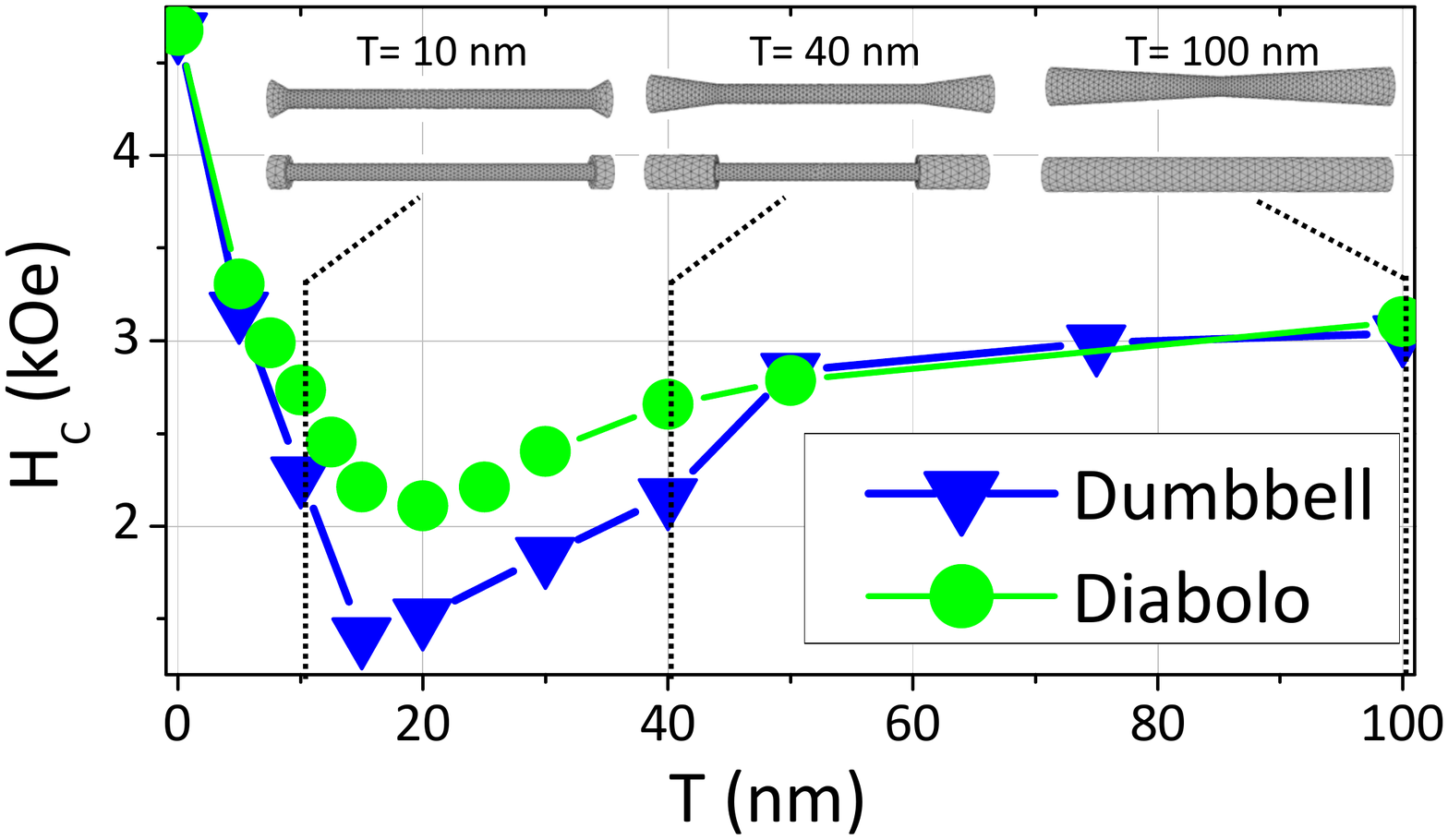}

\caption{Coercive field values for dumbbell and diabolo nanowires with an aspect
ratio $\frac{L}{d}=20$ (i.e. $L=200$ nm, $d=$10 nm and $D=20$
nm) as function of the thickness $T$ of the edge; meshes of the nanowires
for 3 different thickness are shown in the upper part of the figure.}

\label{Fig_T_Dependence-2} 
\end{figure}

Figure \ref{Fig_T_Dependence-2} shows very clearly a meltdown of
the coercive behavior of both the dumbbell and diabolo nanowires during
the increasing of $T$ value. Interestingly a critical thickness around
$T\sim\frac{L}{d}$ is observed for the minimum value of the coercive
field in both the geometry. Once this minimum reached, the coercive
field goes, for both dumbbell and diabolo nanowires, towards the coercive
field value of the cylinder type nanowires having the aspect ratio
of 20 (see Figure \ref{Fig_Length_Dependence}b)). This is intuitively
understood by the tendency of the nanowires shape (both for dumbbell
and diabolo-see Figure \ref{Fig_Length_Dependence}b)) to resemble
more and more, with the increasing of the head thickness $T$, to
a cylinder type nano-object. This result is coherent with previous
ones stressing out the key role of the tips as nucleation points for
the magnetization reversal mechanism (i.e. melting down of the coercivity
as function of the increasing of $T$ value). Even more, the presence
of a critical thickness for the tip above which the coercivity increases
again to reach the cylinder value, highlights the morphology importance
on the shape anisotropy optimization.

In order to study the effect of the morphology on the nanowires behavior
under the application of a magnetic field, we performed the micromagnetic
simulation of the angular dependence of magnetization curves and consequently
of the coercive field $(H_{C})$ and the saturation one $(H_{S})$.
These angular dependence of $H_{C}$ and $H_{S}$ are compared to
the one coming from the well-known relations for a prolate spheroid
object \cite{Bance,Sun2005}: 
\begin{equation}
H_{C}=2\pi M_{S}\left(N_{z}-N_{x}\right)\sin2\varphi_{H}\label{eq:HC_SW_BIS}
\end{equation}
 and 
\begin{eqnarray}
H_{S} & = & \frac{4\pi M_{S}\left(N_{z}-N_{x}\right)}{\left(\sin^{\frac{2}{3}}\varphi_{H}+\cos^{\frac{2}{3}}\varphi_{H}\right)^{\frac{3}{2}}}H_{C}\label{eq:HS_SW_BIS}
\end{eqnarray}

\begin{figure}
\includegraphics[bb=40bp 110bp 660bp 590bp,clip,width=8.5cm]{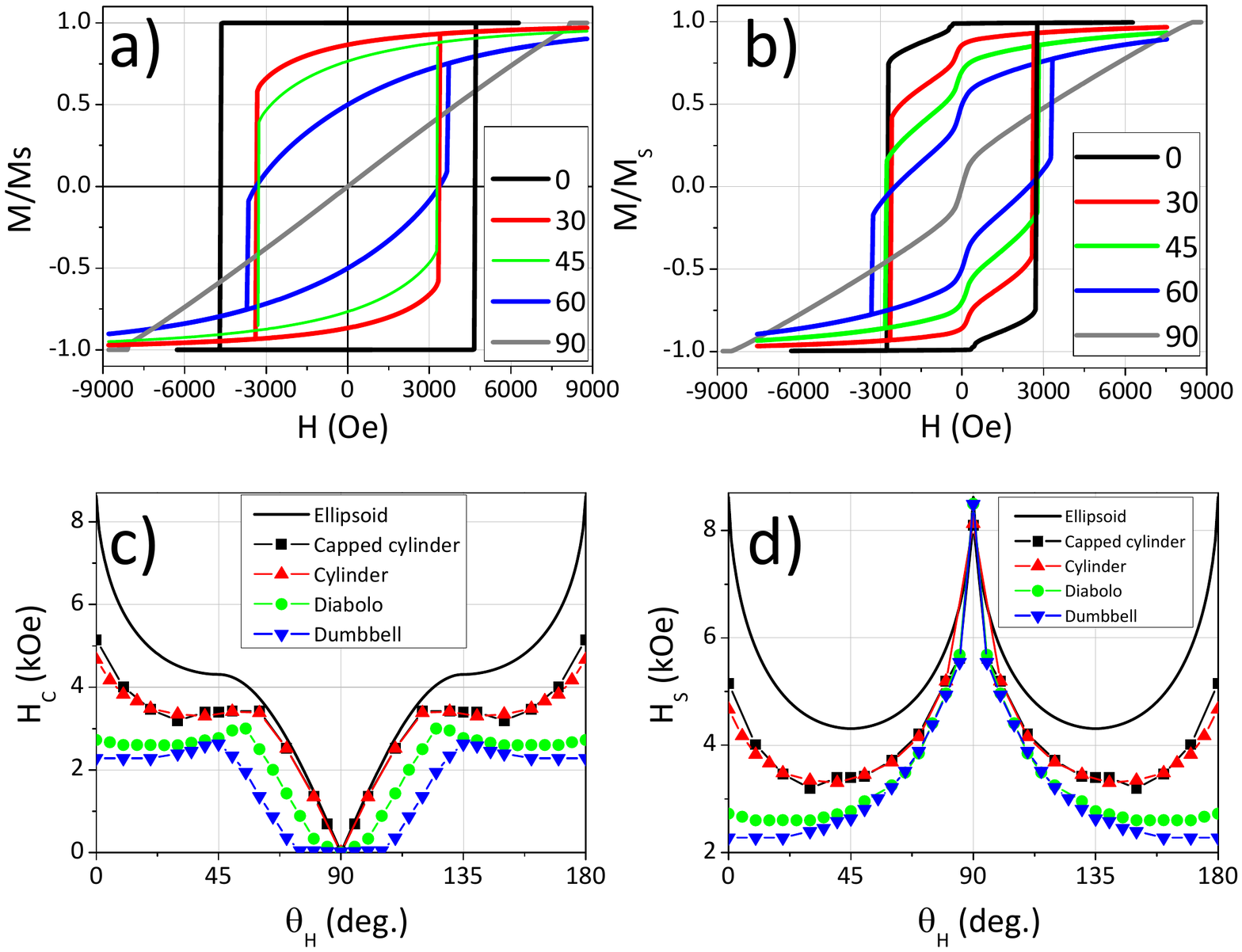}\protect\protect\caption{a) Normalized magnetization curves of a cylinder and b) of a diabolo
for different $\theta_{H}$ angles. c)d) Angular dependence of $H_{C}$
(and $H_{S}$ ) deduced from the magnetization curves for a cylinder
(blue squares), a capped cylinder (green triangles), a diabolo (red
circles) and a dumbbell (black circles). The continuous black lines
in c) and d) correspond to the ellipsoid case and have been calculated
by using equations \ref{eq:HC_SW_BIS} and \ref{eq:HS_SW_BIS}. The
length of these nano-objects is $L=200$ nm while the internal diameter
is $d=10$ nm; in addition, $D$ and $T$ have been fixed to 20 nm
and 10 nm, respectively.}

\label{Fig_HC_Angular} 
\end{figure}

We remained here that $\theta_{H}$ is the angle between the applied
magnetic field and the revolution axis of the nano-objects. In Figure\ref{Fig_HC_Angular}a)
only the cylinder hysteresis normalized loops are presented while
in Figure \ref{Fig_HC_Angular}b) we reported the diabolo ones. Again,
capped cylinder and dumbbell behaviors are not reported because of
their great similarity with respectively the cylinder and the diabolo
shapes. The cylinder (and capped cylinder) type hysteresis loops show
hysteresis behavior very close to the prolate ellipsoid with however
lower coercive field values. For the diabolo (and dumbbell) case,
hysteresis loops are again much complex and put into evidence the
linear dependence before the reversal, characteristics of the non
uniform magnetization distribution promoted by the tip. In Figure
\ref{Fig_HC_Angular}c) and d), we reported the angular dependence
of $H_{C}$ and $H_{S}$ given by equations \ref{eq:HC_SW_BIS} and
\ref{eq:HS_SW_BIS} (for the prolate ellipsoid). It is clear that
for both $H_{C}$ and $H_{S}$, in the case of cylinder type nanowires
(and capped cylinder), the behavior is very close to the prolate ellipsoid
one with a minimum at $\varphi=45^{o}$ (i.e. as expected from the
Stoner-Wohlfarth model). The effect of the morphology (cylinder compared
to the prolate ellipsoid) is given by the lower values of both fields
which was confirmed also in previous works\cite{Brown}. This drop
is, of coarse, more visible when the magnetic field is applied along
the nanowires ($\varphi_{H}=0^{o}$) than when they are close to the
perpendicular direction ($\varphi_{H}=90^{o}$). This is due to the
fact that tip effect is less and less ``efficient'' in the perpendicular
configuration. Indeed, the demagnetizing field at the saturation in
this case, is uniform and its value is closely to the one of an infinite
cylinder with a magnetization at saturation directed perpendicularly
to the nanowire length. For the diabolo (and dumbbell), $H_{C}$ et
$H_{S}$ are very far from the Stoner-Wohlfarth model. Both fields
are much lower than the one from this model, all along the angular
dependence. A ``\textit{plateau-like}'' behavior is observed for
both fields up to a misalignment of 45 degrees. Increasing the misalignment
up to the perpendicular configuration ($\varphi_{H}=90^{o}$), lets
both fields behave closely to the other shapes. This is coherent with
the fact that tip effect becomes less and less efficient approaching
the perpendicular configuration in all different morphologies. All
these results are very interesting as they put into evidence that
even if the coercivity is not optimized in the case of diabolo and
dumbbell nanowires, it is also clear that a higher misalignment is
permitted in case of applications. The ``\textit{plateau-like}''
region allows, in this latter case, a quite constant value of the
coercivity for a few degrees misaligned ensemble of nanowires while
this will not be the case for the cylinder type case (i.e. few degrees
of misalignment will lower the coercivity of few thousands of Oe).

\section{Conclusion}

The static magnetic properties of complex shape Cobalt nanowires have
been studied by micromagnetic simulations. The relevance of each characteristic
size and their critical values strongly confirms that once the aspect
ratio ($\frac{L}{d}$) of the nanowire is bigger than around 10, the
length is no more the pertinent parameter and instead the internal
diameter and the nanowires tips play a key role. As a consequence,
all the experimental effort focusing on improving this aspect ratio
value by extending the length of nanowires at a fixed $d$ will be
useless in term of shape anisotropy optimization. In addition, angular
dependence of magnetization curves reveal that cylinder and capped
cylinder shapes behavior (in term of coercive $H_{C}$ and saturating
fields $H_{S}$) are close to the ``ideal'' ellipsoid case (Stoner-Wolfarth
model) with however lower $H_{C}$ and $H_{S}$ values. In contrary,
diabolo and dumbbell shapes behavior grew apart from this standard
model. We attribute these behaviors to the non uniform demagnetizing
field distribution which is localized inside the nanowire head (for
the whole studied shapes) and which acts as a nucleation point for
the magnetization reversal process. In addition, a clear meltdown
of the coercive behavior for both dumbbell and diabolo nanowires has
been found when increasing the thickness $T$ of the tips. A critical
thickness around $T\sim\frac{L}{d}$ is observed for the minimum value
of the coercive field in both diabolo and dumbbell geometry. Finally,
concerning the potential integration of this kind of nanowires as
building bricks for permanent magnet fabrication, we found that a
very weak angular dispersion of the nanowires is mandatory to keep
a high macroscopic coercive field when using cylinder or capped cylinder
nanowires while in case of diabolo and dumbbell nanowires, the ``\textit{plateau-like}''
behavior allow strong misalignment. 
\begin{acknowledgments}
This work has been partially supported by the Université Paris 13
through a ``Bonus Qualité Recherche'' project and by the Region
Ile-de-France in the framework of C'Nano IdF, the nanoscience competence
center of Paris Region (Eco-Nano project).\end{acknowledgments}

\end{document}